\documentclass[pra,reprint]{revtex4-1}
\usepackage{graphicx}
\usepackage{dcolumn}
\usepackage{bm}
\usepackage{amsmath} 
\usepackage{amsfonts}
\usepackage{psfrag}
\usepackage[pdftex,bookmarks,colorlinks=true,linkcolor = magenta,urlcolor  = red,citecolor = blue]{hyperref}
 \usepackage[usenames,dvipsnames]{color}
 \usepackage{soul}
 \usepackage{ulem}
 \usepackage{cancel}

\def\bra#1{\mathinner{\langle{#1}|}} 
\def\ket#1{\mathinner{|{#1}\rangle}}

\newcommand{\Eq}[1]{Eq. (\ref{#1})}

\newcommand{\ca}[2]{c_{#1,\mathbf{#2}}}
\newcommand{\cc}[2]{c^{\dagger}_{#1, \mathbf{#2}}}

\newcommand{\aaa}[2]{a_{\mathbf{#1},#2}}

\newcommand{\sub}[1]{_{\mathbf{#1}}}

\begin{document}

\title{Theory of intersubband resonance fluorescence}
\author{Nathan Shammah and Simone \surname{De Liberato}}
\affiliation{School of Physics and Astronomy, University of Southampton, Southampton, SO17 1BJ, United Kingdom}
\begin{abstract}

The fluorescence spectrum of a strongly pumped two level system is characterized by the Mollow triplet that has been observed in a variety of systems, ranging from atoms to quantum dots and superconducting qubits. 
We theoretically studied the fluorescence of a strongly pumped intersubband transition in a quantum well.  Our results show that the many-electron nature of such a system leads to a modification of the usual Mollow theory. 
In particular, the intensity of the central peak in the fluorescence spectrum becomes a function of the electron coherence, allowing access to the coherence time of a two dimensional electron gas through a fluorescence intensity measurement. 
\end{abstract}
\maketitle  

\begin{figure}[t!]
\begin{center}
\includegraphics[width=8.6cm]{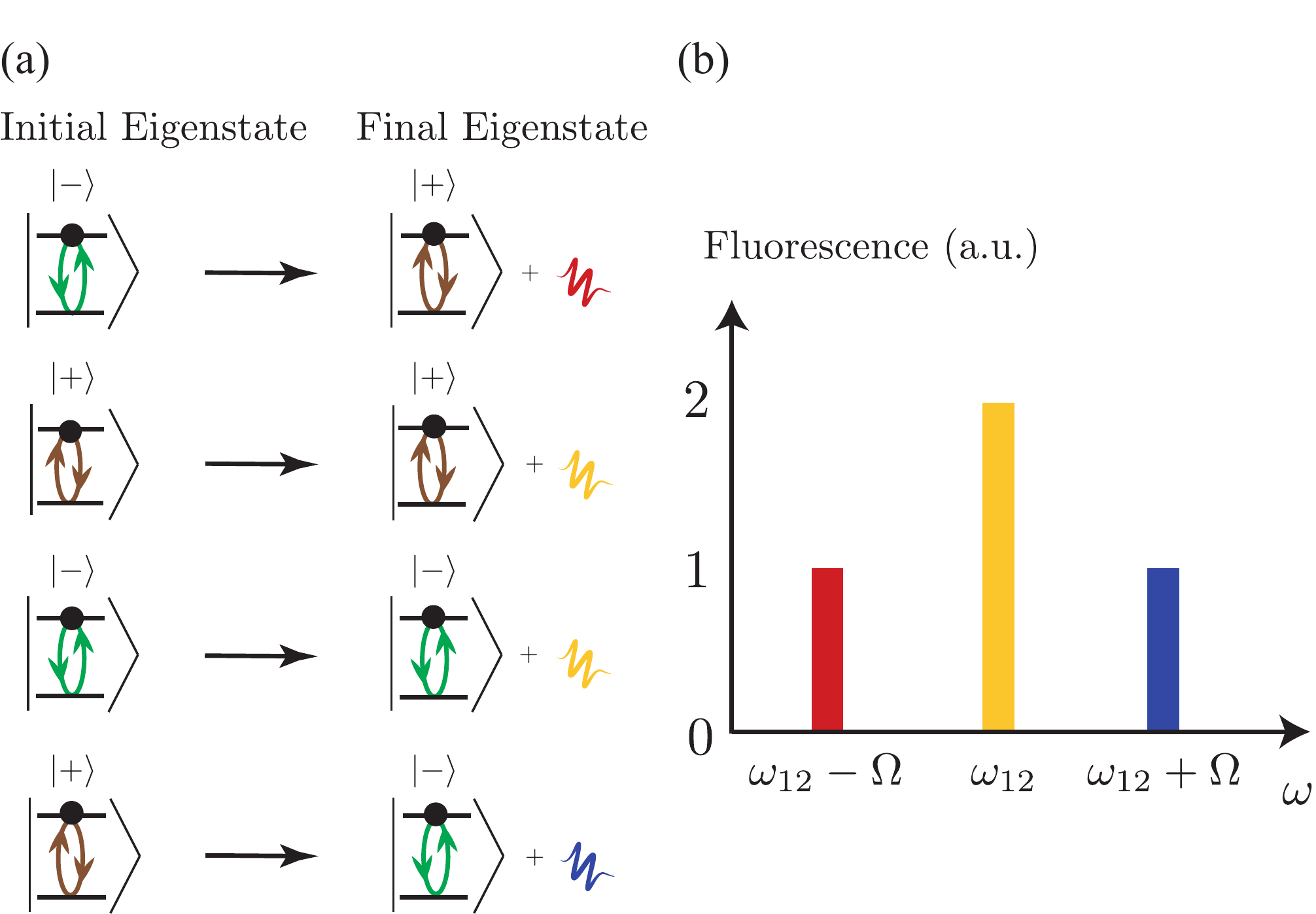}
\caption{\label{fig1}
(a) The transitions between the eigenstates of a strongly illuminated TLS giving rise to the Mollow triplet.
(b) Emission lines from the dressed states occur at the bare frequency $\omega_{12}$ and at the shifted frequencies $\omega_{12}\pm\Omega$.}
\end{center}
\end{figure}

The fluorescence spectrum of a resonantly pumped two level system (TLS) is characterised by the three peak profile named after Mollow \cite{Mollow69}. 
Under strong pumping, the levels of the TLS are split by the ac Stark effect into doublets, $\ket{\pm}$, whose splitting, $\Omega$, is directly proportional to the pump amplitude. Of the four possible spontaneous emission channels between two doublets, two are resonant with the bare frequency of the TLS, $\omega_{12}$, and the other two are at frequency $\omega_{12}\pm\Omega$ [see Fig. \ref{fig1} for a scheme of the emission channels and frequencies]. 
From this simple picture it can be inferred that the emission from the central peak and from the satellites are in ratio 1:2:1, a result that still holds for more refined theoretical approaches \cite{Carmichael76,Ficek99,Delvalle10}.
To date the Mollow triplet has been observed in a wealth of different systems well modelled by a TLS: atoms
\cite{Schuda74,Grove77}, quantum dots \cite{Xu07,Muller07,Vamivakas09,Flagg09,Wei14,He15},
single molecules \cite{Wrigge08}, superconducting qubits \cite{Baur09,Astafiev10,vanLoo13}, and semiconductor quantum well (QW) excitons \cite{Quochi98, Saba00}.

\begin{figure}[t!]
\begin{center}
\includegraphics[width=8.6cm]{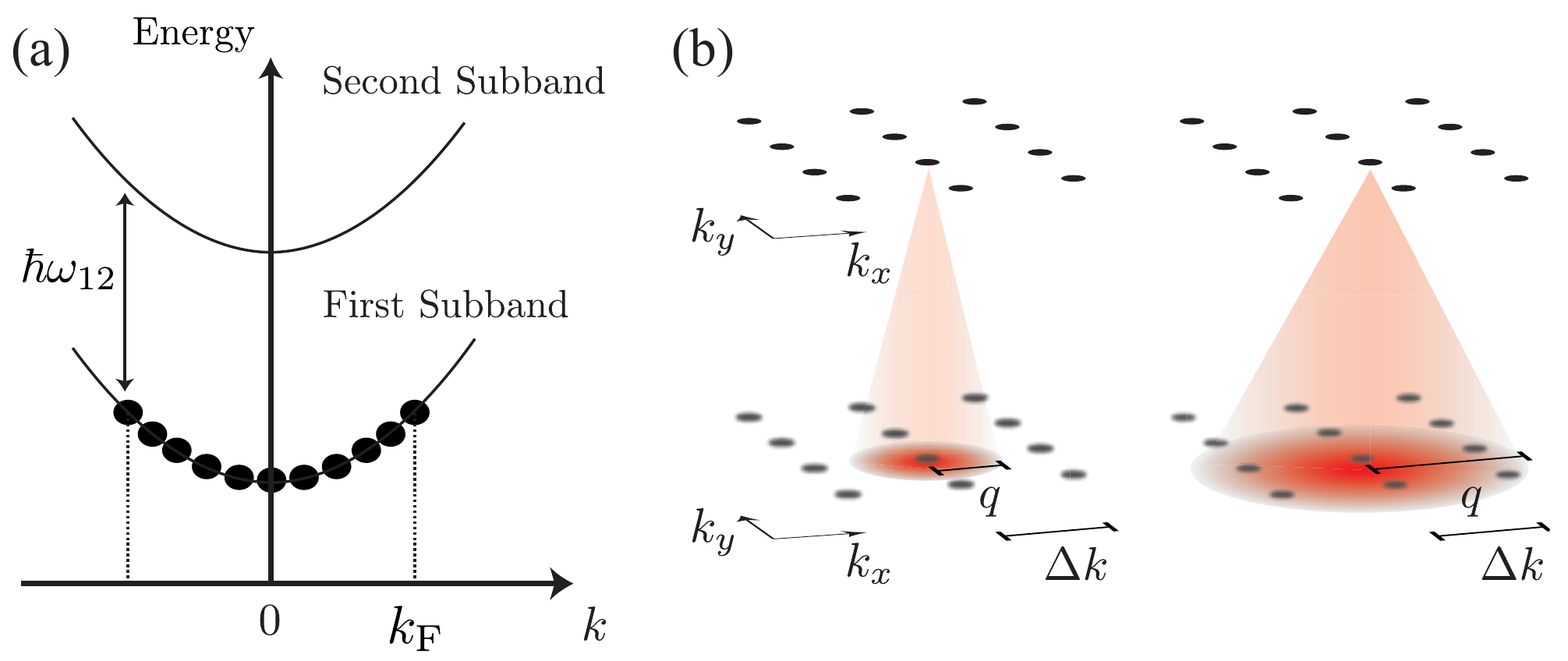}
\caption{\label{fig2}
(a) Energy dispersion of the first two conduction subbands in a doped QW as a function of the in-plane wave vector ${k}$. The first conduction subband is partially filled with a 2DEG up to the Fermi wave vector ${k}_\text{F}$. 
(b) The ratio between the emitted photon wave vector, $q$, and the electron wave vector uncertainty $\Delta k$, determines how many final states are accessible for a photonic emission.}
\end{center}
\end{figure}

Intersubband transitions (ISBTs) occur between two QW conduction subbands, with the lower one containing a two dimensional electron gas (2DEG) created through doping \cite{Helm}, temperature \cite{Anappara07}, or by optically exciting electrons from the valence band \cite{Shtrichman01,Gunter09}. Unbound excitations, ISBTs have a narrow absorption line thanks to the fact that  conduction subbands are quasi-parallel \cite{Nikonov97,Batista04} (see Fig. \ref{fig2} (a) for a graphical representation). Although in ISBTs the Mollow triplet has not yet been directly measured, the ac Stark effect, which can be interpreted as indirect evidence, has been clearly observed \cite{Dynes05}. While ISBTs are usually theoretically described as collections of independent TLSs \cite{Ciuti05, Dynes05, Dynes05b, Frogley06}, it has recently been shown \cite{DeLiberato13} that such an approximation breaks down in the nonlinear regime, when a macroscopic fraction of the total number of electrons is in the excited subband. As explained in Ref. \cite{DeLiberato13}, the origin of such a difference can be intuitively traced to the different dimensionality of the Hilbert space: $2^N$ for a collection of $N$ TLSs and $\left (\begin{array}{c} 2N \\ N \end{array} \right )$ for an ISBT involving $N$ electrons and neglecting border effects, which tends to $\frac{4^N}{\sqrt{\pi N}}$ for large $N$. 

In a setup adapted to measure the resonance fluorescence of the system, almost all the electrons in the first subband are excited at the same time \cite{Dynes05}. 
We can thus expect that the TLS approximation, used to derive the theory of the Mollow triplet, will break down.
In this paper we will develop a predictive many-electron theory describing resonant fluorescence emission from ISBTs and other similar systems beyond the TLS approximation \cite{DeLiberato13,Zhang14}. We show that the physical picture is rather different from the TLS case and that, while the three peaked structure is conserved, the relative intensity of the peaks becomes a function of the coherence of the electron gas. Our work thus hints at the intriguing possibility of measuring the coherence time of a 2DEG through a measure of its resonance fluorescence. In the following we introduce the theoretical description of a resonantly pumped ISBT and develop a perturbative theory of fluorescence emission that will allow us to calculate the intensity of the emitted radiation. We conclude by considering the limitations of our theory and the experimental constraints required to observe such an effect.
The interested reader will find explicit analytic derivations in the Appendix.
\begin{figure*}[t!]
\begin{center}
\includegraphics[width=16cm]{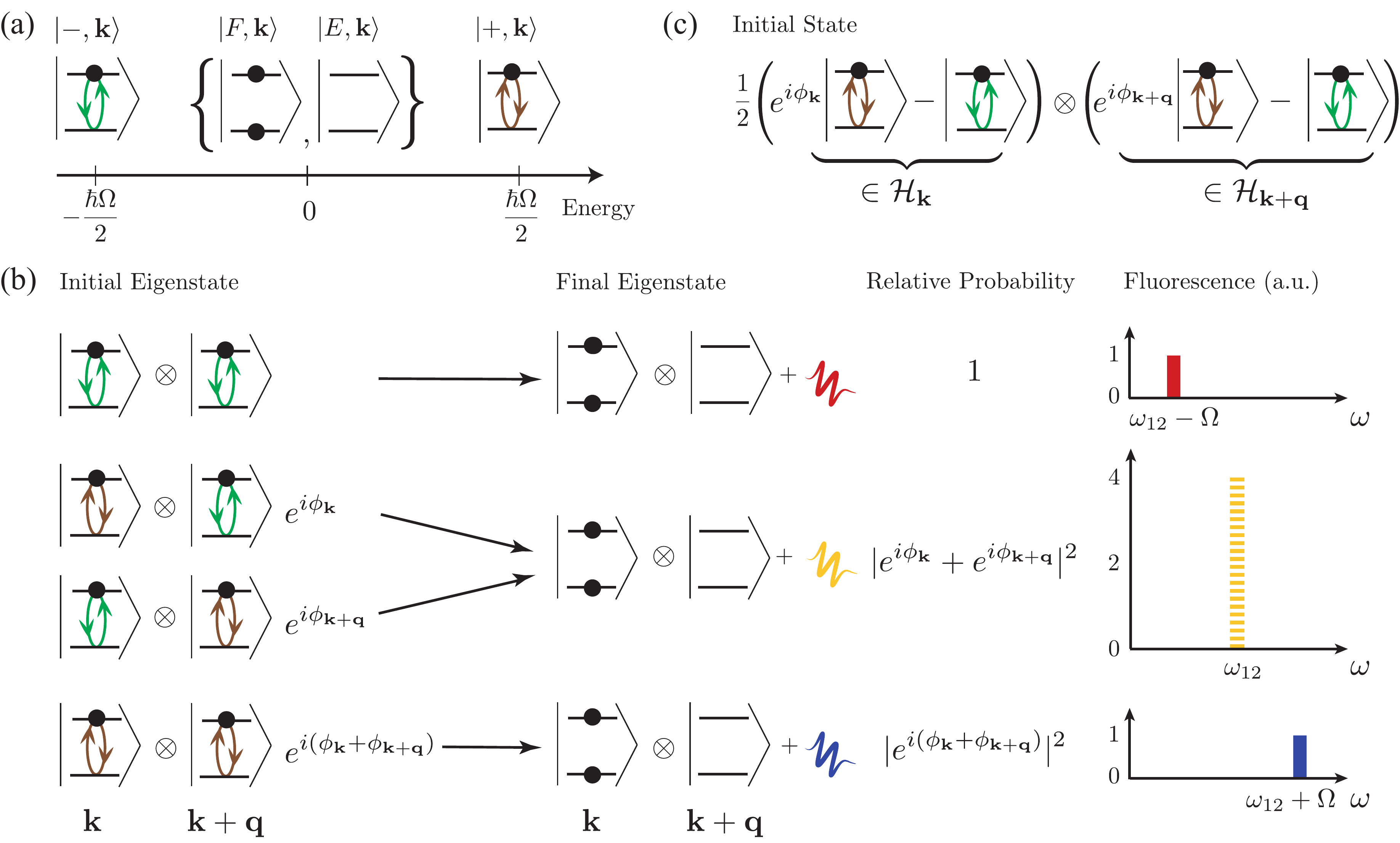}
\caption{\label{fig3}
(a) Eigenvectors and eigenenergies of one of the spaces  for one of the subspaces $\mathcal{H}_{\mathbf{k}}$. 
(b) Transitions in intersubband resonance fluorescence. At $\omega_{{12}}$, two different scattering processes interfere 
and the intensity of the emitted light is given by their relative phases, $4\cos^2\left(\tfrac{\phi_\mathbf{k}-\phi_\mathbf{k+q}}{2}\right)$.  
(c) Two-electron state in which the electrons oscillate between the two subbands with phases $\phi\sub{k}$ and $\phi\sub{k+q}$. }
\end{center}
\end{figure*}

Following the theory developed in Ref. \cite{Shammah14}, we consider two conduction subbands, the lower one containing a 2DEG, pumped by a coherent source resonant at the ISBT transition frequency between the two subbands, $\omega_{12}$ [see Appendix \ref{AppendixA}]. 
Being the in-plane wave vector conserved in a planar structure, the pump, which has a well-defined wave vector $\bar{\mathbf{q}}$, couples each electronic state of the first subband, with wave vector $\mathbf{k}$, to a single state in the second one, with wave vector  $\mathbf{k+\bar{q}}$, and vice versa. The set of all the electron levels in the two subbands is divided into coupled pairs, and the Hilbert space of the electronic system is consequently partitioned into four-dimensional subspaces. 
If only one of each pair of levels is occupied by an electron, then we can consider only two states in each of these subspaces, and the system exactly maps onto a set of pumped TLSs, each one having eigenvalues $\hbar\omega_{\pm, \mathbf{k}}=\pm\frac{\hbar\Omega}{2}$ in the referential frame rotating at the pump frequency, where $\Omega$ is the Rabi frequency proportional to the pump amplitude. The relative eigenstates are
\begin{eqnarray}
\ket{\boldsymbol{\pm},\mathbf{k}}&=&\frac{1}{\sqrt{2}}\left(c^{\dagger}_{2,\mathbf{k+\bar{q}}}\pm c^{\dagger}_{1,\mathbf{k}}\right)\ket{0_\text{el}},
\end{eqnarray}
where $\ket{0_\text{el}}$ describes the empty subbands and $c^{\dagger}_{j,\mathbf{k}}$ is the creation operator for an electron in the $j$th subband, $j=\{1,2\}$, with in-plane wave vector $\mathbf{k}$.
As all interactions are spin-conserving, all sums over electronic wave vectors run implicitly also over spin.
In order to span the whole Hilbert space of the system, we also have to include the states in which neither or both levels of the pair are occupied, $\ket{F,{\mathbf{k}}}=c^{\dagger}_{2,\mathbf{k+\bar{q}}}c^{\dagger}_{1,\mathbf{k}}\ket{0_\text{el}}$ and $\ket{E,{\mathbf{k}}}=\ket{0_\text{el}}$. As these states are either empty or Pauli blocked, they do not couple with the pump, and have eigenvalue zero in the rotating frame \cite{Shammah14,Shammah14b}, as shown in Fig. \ref{fig3} (a).
To each $\mathbf{k}$ value we can thus associate the four-dimensional Hilbert space: $\mathcal{H}_{\mathbf{k}}\equiv \{\ket{-,\mathbf{k}}
,\ket{F,\mathbf{k}},\ket{E,\mathbf{k}},\ket{+,\mathbf{k}} \}.$
After having solved the Hamiltonian for the closed system, in order to describe its fluorescence, we need to couple it to the free electromagnetic field. Calling $a^{\dagger}_{\mathbf{q},q_z}$ the creation operator for a photon of energy $\hbar\omega_{{q},q_z}$, in-plane wave vector $\mathbf{q}$, and out-of-plane wave vector $q_{z}$, the interaction takes the form
\begin{eqnarray}
\label{v}
V&=&\sum_{\mathbf{k},\mathbf{q},q_z} \chi_{{q},q_{z}}c_{2,\mathbf{k+q}}^{\dagger}c_{1,\mathbf{k}}\aaa{q}{q_z}+\text{ H.c.},
\end{eqnarray}
where 
$\chi_{q,q_z}$ is the light-matter coupling term, directly proportional to the ISBT dipole moment. 
As only transverse magnetic modes couple to ISBTs, we neglect photon polarization Ê\cite{Helm}.
Contrary to the TLS case, the interaction in \Eq{v} causes scattering between two-electron eigenvectors, i.e., involving in general states belonging to $\mathcal{H}_{\mathbf{k}}\! \otimes\!\mathcal{H}_{\mathbf{k+q}}$. 
These processes, initially studied in Ref. \cite{Shammah14}, are shown in Fig. \ref{fig3} (b), for the two-electron initial state in Fig. \ref{fig3} (c). 

There are four different scattering channels, two emitting at frequency $\omega_{12}$ and the other two at $\omega_{12}\pm\Omega$. 
While this could seem to lead to a result analogous to that obtained with standard TLS Mollow theory, there is a fundamental difference: here the final state of all the emission channels is the same, which is the product of a fully occupied and an empty state, $\ket{F,\mathbf{k}}\!  \otimes\! \ket{E,\mathbf{k+q}}$. 
This opens up the possibility of having interference for the emission of the central peak, with its intensity depending on the relative phases between the two electrons involved. 
A simple understanding of the phase-dependence of the emission intensity can be gained by considering the semi-classical picture of the two electrons cycling between the two subbands, driven by the pump. The interaction Hamiltonian in \Eq{v} leads to a diagonal process, in which one of the electrons scatters from its own upper state to the other electron's lower state. This transition is Pauli blocked unless both electrons are at the same time in their upper state. The emission intensity will thus depend on the relative phase of the two cycling electrons.  

In order to make quantitative such a handwaving discussion, we need to describe the full many-body state of the system undergoing the Rabi oscillations, taking care of the relative phases of the electrons, $\phi\sub{k}$, which will have a paramount role in the following. Under the approximation that all the electrons participate \cite{Dynes05}, we can write this state as [see Appendix \ref{AppendixB}] 

\begin{eqnarray}
&&\ket{\psi_\text{el}(t)}=\bigotimes\limits\sub{k}\frac{e^{i\phi\sub{k}-i(\Omega/2)t}\ket{+,\mathbf{k}}-e^{i(\Omega/2)t}\ket{-,\mathbf{k}}}{\sqrt{2}}\\
&&=\bigotimes\limits\sub{k}e^{i(\phi\sub{k}/2)} \lbrack i\sin(\tfrac{\phi\sub{k}-\Omega t}{2})c^{\dagger}_{2,\mathbf{k+\bar{q}}}+ 
\cos(\tfrac{\phi\sub{k}-\Omega t}{2})c^{\dagger}_{1,\mathbf{k}}\rbrack\ket{0_\text{el}}.\nonumber
\end{eqnarray}
This state is not stationary and as such it cannot be used in the standard formulation of the Fermi golden rule to calculate the photon emission.
An equivalent formulation can be obtained by calculating, to the first order in  $V$, 
the total number of emitted photons per unit time,
leading to \cite{Lewenstein87}, as shown in Appendix \ref{AppendixD},
\begin{eqnarray}
\Gamma
&=&\frac{1}{ 4\hbar^{2}t}\sum_{\mathbf{k},\mathbf{q},q_{z}}|\chi_{q,q_{z}}\int_{-t/2}^{t/2}
e^{i(\omega_{12}-\omega_{q,q_{z}})\tau}
f_{\mathbf{k},\mathbf{q}}(\tau) d \tau|^{2},
\label{nph1}
\end{eqnarray}
with
$f_{\mathbf{k},\mathbf{q}}(t)=\cos(\tfrac{\phi\sub{k}+\phi\sub{k+q}}{2}-\Omega t)-\cos(\tfrac{\phi\sub{k}-\phi\sub{k+q}}{2})$.
In order to calculate the sums in \Eq{nph1} we neglect specific correlations between the phases, considering them as independent and identically distributed random variables. 
This is a mean-field approximation that allows us to replace the sum over $\mathbf{k}$ with an average over the phase distribution, multiplied by the total number of electrons $N$. 
While an exact investigation of the phase dynamics is beyond the scope of the present work, we can give a qualitative description of phase diffusion. 
In a typical experiment the phases will be initially all equal, as all the electrons lie in the lower conduction subband. Once the continuous drive of the pump is turned on, the electrons will eventually diffuse with a coherence time $\tau_{\text{coh}}$, leading to the experimentally observed wash-out of the Rabi oscillations \cite{Binder90,Schulzgen99,Zrenner02}, see Appendix \ref{AppendixC} for details. Under this approximation, and for times long enough to be able to resolve the peaks ($t>\frac{2\pi}{\Omega}$), we can calculate the emission rate as [see Appendix \ref{AppendixD}]
\begin{eqnarray}
\Gamma
&=&\tfrac{1}{8}\left(1+\langle\cos\phi\rangle^{2}+\langle\sin\phi\rangle^{2}\right)\Gamma_{{0}}(\omega_{12})\nonumber\\
&&+\tfrac{1}{16}\Gamma_{{0}}(\omega_{12}+\Omega)+\tfrac{1}{16}\Gamma_{{0}}(\omega_{12}-\Omega)
 \label{G1},
\end{eqnarray}  
where $\Gamma_{{0}}(\omega)$ is the standard dipole emission at frequency $\omega$ and the angular bracket is the average over the phase $\phi$. While electrons are fully coherent, the phase distribution will be strongly peaked around a well defined value $\bar{\phi}$, and in \Eq{G1} we obtain $\langle\cos\bar{\phi}\rangle^{2}+\langle\sin\bar{\phi}\rangle^{2}=1$, independently of the specific value of $\bar{\phi}$.
If we approximate $\Gamma_{{0}}(\omega_{12}\pm\Omega)\simeq\Gamma_{{0}}(\omega_{12})$, the intensity of the central peak is in a ratio 1:4:1 with respect to the satellites. For $t\gg\tau_{\text{coh}}$ instead, the electron coherence is lost. The phases can in this case be considered as uniformly distributed and we have $\langle\cos\phi\rangle=\langle\sin\phi\rangle=0$, recovering the usual ratio 1:2:1 for the area of the three peaks and the expected asymptotic value for the total incoherent emission rate $\Gamma_{\text{in}}=\frac{\Gamma_{{0}}}{4}$ \cite{Shammah14} [see Appendix \ref{AppendixD}].

\begin{figure}[t!]
\begin{center}
\includegraphics[width=8.6cm]{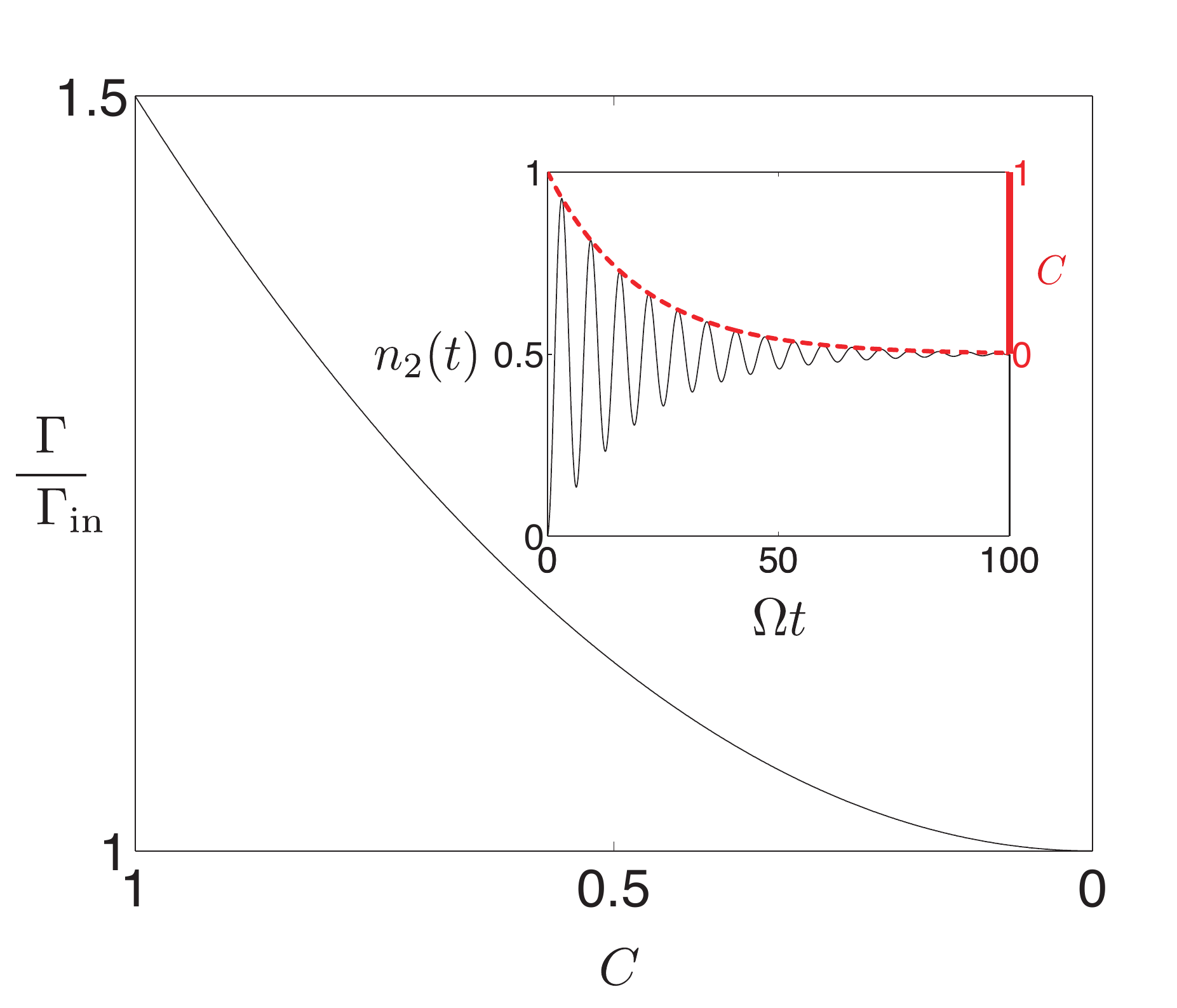}
\caption{\label{fig4}
The integrated intensity of fluorescence, plotted normalized to the fully incoherent value $\Gamma_{\text{in}}$, decreases as the visibility $C$ goes to zero. 
Inset: The collective Rabi oscillations of the second subband, $n_{2}(t)$ (black solid line, left axis) are washed out in a time $\tau_{\text{coh}}$. The visibility $C$ (red dashed line, right axis) decreases accordingly in time. Here $\Omega\tau_{\text{coh}}=20$. }
\end{center}
\end{figure}

A clear way to present this result is in terms of the many-electron Rabi oscillations. The fraction of electrons in the second subband can be calculated as [see Appendix \ref{AppendixC}]
\begin{eqnarray}
\label{n2t}
n_2(t)&=&\bra{\psi_\text{el}(t)}\tfrac{\sum_{\mathbf{k}}c_{2,\mathbf{k}}^{\dagger}c_{2,\mathbf{k}}}{N}
\ket{\psi_\text{el}(t)}=\langle \sin^2(\tfrac{\phi-\Omega t}{2}) \rangle,
\end{eqnarray}
which have a visibility decreasing exponentially in time due to decoherence (see the inset of Fig. \ref{fig4}), given by an envelope function normalized between $0$ and $1$,
\begin{eqnarray}
C&=&2\max_t n_2(t)-1=\sqrt{\langle \cos\phi\rangle^2 +\langle \sin\phi\rangle^2}.
\label{C}
\end{eqnarray}
Remarkably, \Eq{G1} can be recast as a simple function of $C$,   
\begin{eqnarray}
\Gamma&=&\tfrac{1}{8}(1+C^{2})\Gamma_{{0}}(\omega_{12})+\tfrac{1}{16}\Gamma_{{0}}(\omega_{12}+\Omega)+\tfrac{1}{16}\Gamma_{{0}}(\omega_{12}-\Omega),\quad\,\,
\label{G2}
\end{eqnarray}
which is plotted in Fig. \ref{fig4}, assuming that $\Gamma_{{0}}(\omega_{12}\pm\Omega)\simeq\Gamma_{{0}}(\omega_{12})$, valid for $\omega_{12}\gg\Omega$. From \Eq{G2} we can clearly see how the relative intensity of the central peak decreases from $4$ to $2$ with dephasing.

Notice that, while the coherence time can be measured by four-wave mixing \cite{Kaindl98}, or by the populations, through \Eq{n2t} \cite{Zrenner02}, this requires one to perform either a pump and probe measurement, or to have an appositely crafted system with a shelving level  \cite{Dynes05}. 
Through $\Gamma$ instead, the same information can be acquired via a time-resolved fluorescence measurement. A main experimental challenge will be to discriminate the weak fluorescence signal from the pump. While polarization cannot be used to this aim, due to ISBTs selection rules, it is possible to exploit the broad angular distribution of the fluorescence. Similarly to what was done in Ref. \cite{Muller07}, the sample can be engineered for the pump beam to be confined in a waveguided mode, while part of the fluorescence is emitted in non-guided modes, and collected outside of the sample.

Above we implicitly assumed that the visibility $C$ can be considered almost constant over a few Rabi oscillations, in order to neglect the time dependency of \Eq{C} in \Eq{nph1}.
Such a condition, which can be expressed as ${\Omega}\tau_{\text{coh}}\gg1$, can be fulfilled by present-day technology.
The main contributions to $\tau_{\text{coh}}$ generally come from interface roughness scattering (IRS) and longitudinal optical phonon emission, which are both suppressed in wide QWs  \cite{Ferreira89,Helm,Kaindl01,Unuma03,Virgilio14}. Since IRS scales as the inverse sixth power of the QW length and the phonon emission is forbidden when the intersubband gap becomes smaller than the optical phonon frequency \cite{Murdin97}, coherence times of the order of a picosecond are achievable in wide QWs.
In particular, in GaAs/AlGaAs QWs similar to the ones investigated in Ref. \cite{Campman96}, of bare frequency $\hbar\omega_{12}=100$ meV, at $T=4.2$ K, linewidths of $2.5$ meV, corresponding to $\tau_{\text{coh}}=0.5$ ps, can be obtained  even under intense pumping \cite{Dynes05,Frogley06,Murphy14}.  
In such structures, Rabi splittings in excess of $10$ meV are achievable for internal pump intensities of the order of $\sim $MW/cm$^{2}$ \cite{Kaindl01, Dynes05}, leading to ${\Omega}\tau_{\text{coh}}\simeq 10$. 
In Ge/SiGe QWs, the non-polarity of the Ge lattice inhibits Fr\"{o}hlich-mediated phonon emission and removes the constraint of performing measurements at cryogenic temperatures.
While no observation of Rabi oscillations has been made in these systems, recent results with linewidths of a few meV have been obtained up to $T=300$ K for the bare frequencies, so $\tau_{\text{coh}}=1$ ps or longer is reachable by current technology \cite{Virgilio14b}. 
For a structure with $\hbar\omega_{12}=50$ meV and $L_{QW}=10$ nm as in Ref. \cite{Virgilio14b}, we expect ${\Omega}\tau_{\text{coh}}\simeq 20$. 

We developed our theory neglecting the Coulomb interaction. While it is well known that in the case of parallel subbands its effects usually reduce to a renormalization of the intersubband transition energy \cite{Nikonov97} (the so-called depolarization shift, vanishing for parabolic wells \cite{Geiser12,Kohn61}), Coulomb interaction could have a non-negligible impact in our case, as the presence of collective excitations \cite{Todorov10,DeLiberato12,Delteil12,Kyriienko12,Luc13} could spoil the independent electron picture we used \cite{Luo04}. 
For this reason we can consider our analysis as rigorous only in the limit in which the depolarization shift is much smaller than the Rabi frequency, and plasmonic effects can be ignored. In both structures from Refs. \cite{Campman96} and \cite{Virgilio14b} we estimated  a depolarization shift of less than $0.8$ meV for a carrier concentration of $n_{2DEG}=10^{11}$ cm$^{-2}$ \cite{Helm, Busby10,Graf00,Shtrichman01}. Our hypothesis of a depolarization shift much smaller than the Rabi frequency is thus fulfilled for all but the most heavily doped structures.

A last point worth stressing is that we considered a planar infinite 2DEG. 
While this is usually a very good approximation for high mobility samples, we still have to consider that the oscillating electrons are localized in the laser spot. 
If we call $\Delta k$ the electron wave vector uncertainty due to such a confinement, the observation of interference of multi-electron scattering will require that the in-plane momentum of the emitted photon, $q$, obeys the relation $q\gg\Delta k$ [see Fig. \ref{fig2} (b) for a scheme of the emission process]. 
For a typical waist $w\simeq 50$ $\mu$m \cite{Dynes05}, this condition is well fulfilled already for $\hbar\omega_{12}>7$ meV, but attention should be paid when applying our theory to narrow waists and THz QWs.

In conclusion, we have shown that when ISBTs are strongly driven at resonance by an optical pump, the fluorescence has peculiar features that are not found in that of a collection of non-interacting TLSs. 
Although the TLS approximation is useful to describe Rabi oscillations, the richer dynamics introduced by incoherent emission processes requires a more sophisticated approach. 
In ISBTs the ratio between the three resonance fluorescence peaks deviates from the Mollow triplet due to interference effects that enhance the emission from the central peak.
Thanks to this effect, the coherence time of a 2DEG could be accessed directly from a measure of the time resolved fluorescence intensity. 

We thank Alexey Kavokin, Chris Phillips, James Mayoh, Luca Sapienza, and Denis Villemonais for fruitful discussions.
SDL is Royal Society Research Fellow. 
SDL and NS acknowledge support from EPSRC Grant No. EP/L020335/1.

\appendix
\label{Appendix}
\section{Driven Hamiltonian}
\label{AppendixA}

In this Apendix we give a brief review of the theoretical framework, introduced in Ref. \cite{Shammah14}, which we use to study a strongly pumped intersubband transition (ISBT).
The Hamiltonian of the pumped electronic system can be written as
\begin{eqnarray}
\label{h}
H&=&\sum_{j,\mathbf{k}} \hbar\omega_{j,{k}}\cc{j}{k}\ca{j}{k},\nonumber\\
&&+\frac{\hbar\Omega}{2}\sum_{\mathbf{k}}\left( \cc{2}{k+\bar{q}}\ca{1}{k}e^{-i\omega_\text{L} t}+ \text{ H.c.}\right),
\end{eqnarray}
where $\cc{j}{k}$ is the second quantization operator creating an electron in subband $j=\{1,2\}$ with in-plane wave vector $\mathbf{k}$ and energy $\hbar\omega_{j,k}$, $\Omega$  is the Rabi frequency proportional to the pump field amplitude, and $\omega_\text{L}$ is the pump frequency, whose in-plane wave vector is $\mathbf{\bar{q}}$. 
We neglect to explicitly mark the spin index as all the interactions are spin conserving.
In the following all the sums over electronic wave vector, which run until the Fermi wave vector, will thus implicitly sum also over spin.  
Being the photonic wave vector much smaller than the typical electronic one, and given that the conduction subbands are to a good approximation parallel, we can consider $\omega_{12}\simeq \omega_{2,\lvert \mathbf{k+\bar{q}}\lvert}-\omega_{1,k}$. Passing in the frame rotating at the pump frequency, in the resonant case $\omega_\text{L}=\omega_{12}$,
we can reduce the Hamiltonian to the time-independent form
\begin{eqnarray}
H'&=&\frac{\hbar\Omega}{2}\sum_{\mathbf{k}}\left( \cc{2}{\mathbf{k+\bar{q}}}\ca{1}{\mathbf{k}}+ \text{ H.c.}\right).
\label{H'}
\end{eqnarray}
Each value of $\mathbf{k}$ in the sum in \Eq{H'} spans an independent four-dimensional Hilbert subspace, whose eigenvectors are 
\begin{eqnarray}
\ket{\pm,\mathbf{k}}&=&\tfrac{1}{\sqrt{2}}(c^{\dagger}_{2,\mathbf{k}+\bar{\mathbf{q}}}\pm c^{\dagger}_{1,\mathbf{k}})\ket{0_\text{el}},\nonumber \\
\ket{F,\mathbf{k}} &=&c^{\dagger}_{2,\mathbf{k}+\bar{\mathbf{q}}}c^{\dagger}_{1,\mathbf{k}}\ket{0_\text{el}},\nonumber\\
\ket{E,\mathbf{k}} &=&\ket{0_\text{el}},
\label{eigenvectors}
\end{eqnarray}
the first two with energy $\pm\tfrac{\hbar\Omega}{2}$ and the last two with zero energy in the rotating frame. In \Eq{eigenvectors} the ket $\ket{0_\text{el}}$ represents the empty conduction band. 

\section{State of the driven system}
\label{AppendixB}

We assume that initially the pump is ``off'' and that the ground state of the system
thus describes a two dimensional electron gas in the lower subband
\begin{eqnarray}
\label{statenop}
\ket{\psi_\text{el}(0)}=\bigotimes_{\mathbf{k}}c^{\dagger}_{1,\mathbf{k}}\ket{0_\text{el}}.
\end{eqnarray} 
The state in \Eq{statenop} can be rewritten in terms of the two eigenstates of $H'$ in \Eq{eigenvectors} as  
\begin{eqnarray}
\ket{\psi_\text{el}(0)}&=&\bigotimes_{\mathbf{k}}\tfrac{1}{\sqrt{2}}\left(e^{i\phi\sub{k}}\ket{+,\mathbf{k}}-\ket{-,\mathbf{k}}\right),
\label{state}
\end{eqnarray} 
where we added the relative phase of each oscillating electron $\phi_{\mathbf{k}}$. Those phases are all zero at $t=0$, in order to recover \Eq{statenop}, but they can evolve in time due to dephasing processes, which cannot be accounted for in the Hamiltonian dynamics we considered.
When the optical pump is ``switched on'' the electrons start oscillating under the action of $H'$. The state in \Eq{state} can be evolved using \Eq{H'} and \Eq{eigenvectors} as
\begin{eqnarray}
\ket{\psi_\text{el}(t)}&=&e^{-iH't/\hbar}\ket{\psi}\nonumber\\
&=&\bigotimes_{\mathbf{k}}\tfrac{1}{\sqrt{2}}\left(e^{i(\phi\sub{k}-\tfrac{\Omega}{2}t)}\ket{+,\mathbf{k}}-e^{i\tfrac{\Omega}{2}t}\ket{-,\mathbf{k}}\right)\nonumber\\
&=&\bigotimes\limits\sub{k}e^{i\frac{\phi\sub{k}}{2}} \lbrack i\sin(\tfrac{\phi\sub{k}-\Omega t}{2})c^{\dagger}_{2,\mathbf{k+\bar{q}}}+ 
\cos(\tfrac{\phi\sub{k}-\Omega t}{2})c^{\dagger}_{1,\mathbf{k}}\rbrack\ket{0_\text{el}},\nonumber\\
\label{stateevel}
\end{eqnarray}
where in the second line of \Eq{stateevel} we expressed the state in terms of the fermionic operators acting on the empty subbands, which makes it evident that the electrons are oscillating between the two subbands. 
\section{Visibility of the collective Rabi oscillations}
\label{AppendixC}
The fraction of electrons in the second subband can be calculated from the evolved electronic state in \Eq{stateevel} as
\begin{eqnarray}
n_2(t)&=&\tfrac{1}{N}\bra{\psi_\text{el}(t)}\sum_{\mathbf{k}}c_{2,\mathbf{k}}^{\dagger}c_{2,\mathbf{k}}\ket{\psi_\text{el}(t)}\nonumber\\
&=&\tfrac{1}{N}\sum_{\mathbf{k}}\sin^{2}\left(\tfrac{\phi_{\mathbf{k}}-\Omega t}{2}\right)\nonumber\stackrel{i.i.d.}=\langle\sin^{2}\left(\tfrac{\phi-\Omega t}{2}\right)\rangle,\\
\label{nel1}
\end{eqnarray}
where $N$ is the total number of electrons. In the last passage of \Eq{nel1} we assumed that the phases $\phi_{\mathbf{k}}$ are independent identically distributed (i.i.d.) random variables, so that $\tfrac{1}{N}\sum\sub{k}g(\phi\sub{k})=\langle g(\phi) \rangle$ for any function $g(\phi)$.
\begin{figure*}[t!]
\begin{center}
\includegraphics[width=16.5cm]{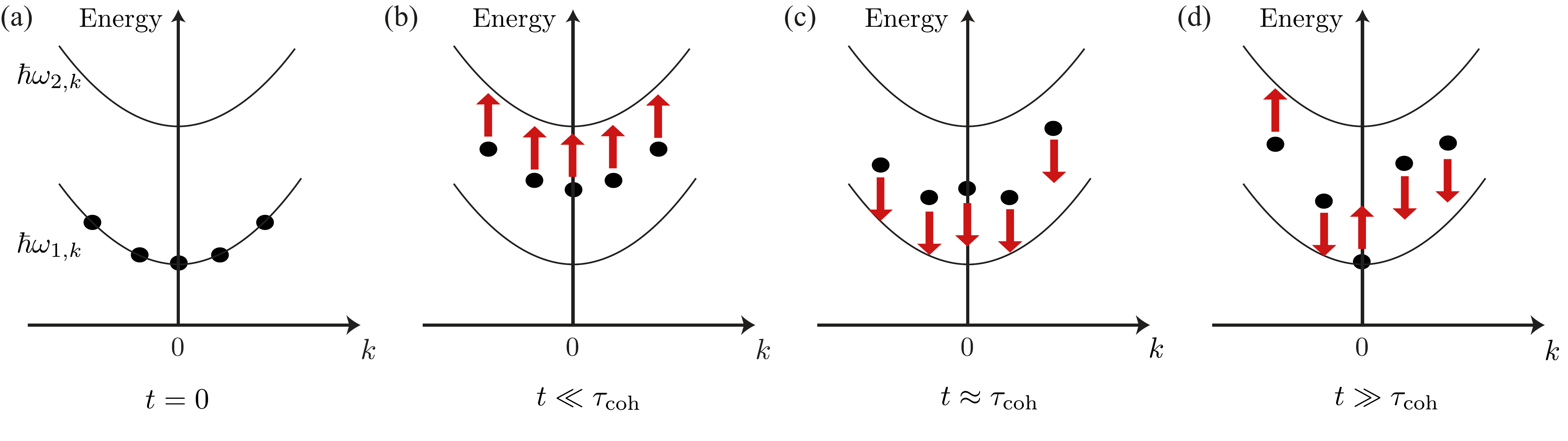}
\caption{\label{figs1}
An illustration of phase diffusion over time. (a) At $t=0$, all electrons lie in the first subband and the state is well described by \Eq{statenop}, that is \Eq{state} with all $\phi\sub{k}=0$.
(b) For $t\ll\tau_\text{coh}$ the electrons oscillate coherently. (c) For $t\approx\tau_\text{coh}$ the electrons start dephasing. 
(d) At $t\gg\tau_\text{coh}$ coherence between electrons is lost and Rabi oscillations are averaged out.}
\end{center}
\end{figure*}
Assuming that Rabi oscillations are faster than dephasing, their visibility can be operationally quantified as the difference between the maximum and the average of the population oscillations, that is
\begin{eqnarray}
C&=&2\left(\max_{t}n_{2}(t) -\tfrac{1}{2}\right),
\label{c}
\end{eqnarray} 
where we chose the normalisation factor in order to have $0\leq C\leq1$.
Inserting \Eq{nel1} in \Eq{c} and developing the sine we obtain
\begin{eqnarray}
C&=&2\max_{t}[\langle\sin^{2}\left(\tfrac{\phi-\Omega t}{2}\right)\rangle] -1\nonumber\\
&=&2\max_{t}[-(\langle\cos^{2}\tfrac{\phi}{2}\rangle\cos^{2}\tfrac{\Omega t}{2}+\langle\sin^{2}\tfrac{\phi}{2}\rangle\sin^{2}\tfrac{\Omega t}{2}\nonumber\\
&&+\tfrac{1}{2}\langle\sin\phi\rangle\sin\Omega t)] +1.
\label{ciid}
\end{eqnarray} 
The maximum over time can be found by taking the time derivative of the expression between the square brackets in \Eq{ciid} and by imposing it equal to zero, that is  
\begin{eqnarray}
\langle\sin\phi\rangle\cos\Omega t+\left(\langle\sin^{2}\tfrac{\phi}{2}\rangle-\langle\cos^{2}\tfrac{\phi}{2}\rangle\right)\sin\Omega t=0,
\label{derz}
\end{eqnarray}
which is satisfied for 
\begin{eqnarray}
\Omega t=\arctan\left(\frac{\langle\sin\phi\rangle}{\langle\cos^{2}\tfrac{\phi}{2}\rangle-\langle\sin^{2}\tfrac{\phi}{2}\rangle}\right).
\end{eqnarray}
After some straightforward algebra \Eq{ciid} can thus be rewritten in its final form as 
\begin{eqnarray}
C&=&\sqrt{\langle \cos\phi\rangle^2 +\langle \sin\phi\rangle^2}.
\label{cfinal}
\end{eqnarray}
Let us discuss qualitatively how the electron coherence of the system changes over time in a typical experiment of continuous pumping. 
At $t=0$ all electrons lie in the first subband and we can assume that all $\phi\sub{k}=0$, as shown graphically in Fig. \ref{figs1}(a). 
As the pump is switched on, the electrons start oscillating and for $t\ll\tau_\text{coh}$ relative phases will not have diffused much, as shown in Fig. \ref{figs1}(b). 
Their distribution will thus be strongly peaked around a well defined value $\bar{\phi}$, and the average values in \Eq{cfinal} will effectively reduce to evaluate the functions at $\bar{\phi}$: $C=\sqrt{\cos^2\bar{\phi} +\sin^2\bar{\phi}}=1$, for any value of $\bar{\phi}$.
 At times comparable to the coherence time, $t\approx\tau_\text{coh}$, with the pump still illuminating the system, more electrons will have been involved in incoherent scattering processes, as visible in Fig. \ref{figs1}(c), decreasing the visibility of the Rabi oscillations.
Finally, for times much longer than the coherence time, $t\gg\tau_\text{coh}$, the phases of the different electrons will be completely randomized, as schematized by Fig. \ref{figs1}(d). 
In this case the phase averages in \Eq{cfinal} will be over an uniform distribution,
$\langle \cos\phi\rangle=\langle \sin\phi\rangle=0$, leading to $C=0$.
\section{Photon emission rate}
\label{AppendixD}
In order to describe the fluorescence we need to introduce the electromagnetic field, whose 
free Hamiltonian can be generally written as
\begin{eqnarray}
H_{\text{ph}}&=&\sum_{\mathbf{q}, q_{z}} \hbar\omega_{{q}, q_{z}}a^{\dagger}_{\mathbf{q}, q_{z}}a_{\mathbf{q}, q_{z}},
\end{eqnarray}
where $a^{\dagger}_{\mathbf{q}, q_{z}}$ is the creation operator of a photon with wave vector $\mathbf{q}$ in the plane of the quantum well and $q_z$ normal to it, whose energy is $\hbar\omega_{{q}, q_{z}}$.
The coupling between the electromagnetic field and the ISBT is given in the rotating wave approximation by the interaction term
\begin{eqnarray}
V&=&\sum_{\mathbf{k},\mathbf{q},q_z} \chi_{{q},q_{z}}c_{2,\mathbf{k+q}}^{\dagger}c_{1,\mathbf{k}}\aaa{q}{q_z}+\text{ H.c.},
\label{v0}
\end{eqnarray} 
that in the rotating frame, and in interaction representation with respect to $H_{\text{ph}}+H'$, becomes
\begin{eqnarray}
V(t)&=&\sum_{\mathbf{k},\mathbf{q},q_z} \chi_{{q},q_{z}}e^{i(\omega_{12}-\omega_{{{q},q_{z}}})t}e^{iH't/\hbar}c_{2,\mathbf{k+q}}^{\dagger}c_{1,\mathbf{k}}\aaa{q}{q_z}e^{-iH't/\hbar}\nonumber\\
+\text{ H.c.}
\label{v1}
\end{eqnarray} 
The rate of the emitted photons from the system, $\Gamma$, can be calculated by dividing the total number of emitted photons by the interaction time 
 \begin{eqnarray}
 \Gamma&=&\frac{N_\text{ph}(t)}{t}=\frac{1}{t}\bra{\psi(t)}\sum_{\mathbf{q},q_z}a^{\dagger}_{\mathbf{q},q_{z}}a_{\mathbf{q},q_{z}}\ket{\psi(t)},
\label{nph1}
\end{eqnarray}
where $\ket{\psi(t)}$, the total system wave vector in interaction representation, can be obtained by calculating to first order in $V$ the evolution of the electronic state in \Eq{state} and of the photonic vacuum, $\ket{0_\text{ph}}$
\begin{eqnarray}
\ket{\psi(t)}&=&\ket{\psi_\text{el}(0)}\!\otimes\! \ket{0_\text{ph}}-\frac{i}{\hbar}\int_{-t/2}^{t/2}V(\tau)\ket{\psi_\text{el}(0)}\!\otimes\! \ket{0_\text{ph}}d\tau.\nonumber\\
\label{stateev}
\end{eqnarray}
Using \Eq{v1} and \Eq{stateev}, the photon population can be calculated as 
\begin{eqnarray}
N_\text{ph}(t)&=&\frac{1}{\hbar^{2}}\iint_{-\frac{t}{2}}^{\frac{t}{2}}d\tau d\tau'\bra{\psi(0)}V(\tau)\sum_{\mathbf{q},q_z}a^{\dagger}_{\mathbf{q},q_{z}}a_{\mathbf{q},q_{z}}V(\tau')\ket{\psi(0)},\nonumber\\
\label{nph2}
\end{eqnarray}and after some straightforward algebra \Eq{nph1} can be rewritten explicitly as   
\begin{eqnarray}
 \Gamma
 &=&\frac{1}{4 \hbar^{2}t}\sum_{\mathbf{k},\mathbf{q},q_{z}}|\chi_{q,q_{z}}\int_{-t/2}^{t/2}e^{i(\omega_{12}-\omega_{q,q_{z}})\tau}f_{\mathbf{k,q}}(\tau)d \tau|^{2},\nonumber\\
\label{nph3}
\end{eqnarray} 
with $f_{\mathbf{k,q}}(t)=\cos(\tfrac{\phi\sub{k}+\phi\sub{k+q}}{2}-\Omega t)-\cos(\tfrac{\phi\sub{k}-\phi\sub{k+q}}{2})$.
The integrand in \Eq{nph3} contains only phase factors linear in $\tau$ and as such,
for long enough times, \Eq{nph3} will be given by a sum of squared Dirac delta-like functions. Using the usual trick commonly employed to derive the Fermi golden rule of formally transforming $\delta(\omega)^2=\delta(0)\delta(\omega)=\frac{t\delta(\omega)}{2\pi}$, we obtain  
\begin{eqnarray}
\Gamma&=&\frac{\pi}{8 \hbar^{2}}\sum_{\mathbf{k,q},q_{z}}|\chi_{q,q_{z}}|^{2} \lbrack4\cos^{2}(\tfrac{\phi_{\mathbf{k}}-\phi_{\mathbf{k+q}} }{2})\delta(\omega_{q,q_{z}}-\omega_{12})\nonumber\\
&&+\delta(\omega_{q,q_{z}}-(\omega_{12}+\Omega))+\delta(\omega_{q,q_{z}}-(\omega_{12}-\Omega))\rbrack.\nonumber\\
\label{dnph}
\end{eqnarray}
As done previously for the electron visibility, we assume that the phases are i.i.d. random variables. The sum over $\mathbf{k}$ of the only term depending on $\mathbf{k}$ in \Eq{dnph} thus becomes
\begin{eqnarray}
&&\frac{1}{N}\sum_{\mathbf{k}} 4\cos^{2}(\tfrac{\phi_{\mathbf{k}}-\phi_{\mathbf{k+q}} }{2})\nonumber\\
&=&\frac{1}{N}\sum_{\mathbf{k}} 
2\left(1+\cos\phi_{\mathbf{k}} \cos\phi_{\mathbf{k+q}} +\sin\phi_{\mathbf{k}}\sin\phi_{\mathbf{k+q}}\right)\nonumber\\
&\stackrel{i.i.d.}=&2(1+\langle\cos\phi\rangle^{2}+\langle\sin\phi\rangle^{2}).
\label{iid}
\end{eqnarray}
Finally, inserting \Eq{iid} into \Eq{dnph}, and exploiting \Eq{cfinal}, we obtain
\begin{eqnarray}
\Gamma&=&
\tfrac{1}{8}(1+C^{2})\Gamma_0(\omega_{12})+\tfrac{1}{16}\Gamma_0(\omega_{12}+\Omega)+\tfrac{1}{16}\Gamma_0(\omega_{12}-\Omega),\nonumber\\
\label{g}
\end{eqnarray}
with $\Gamma_{0}(\omega)$ being the spontaneous emission rate of $N$ two-level systems with bare frequency $\omega$. Notice that, for long times $t\gg\tau_\text{coh}$,
$C=0$ and we recover $\Gamma\simeq\tfrac{\Gamma_0(\omega_{12})}{4}$ where, as discussed in Ref. \cite{Shammah14}, the factor $\tfrac{1}{4}$ comes from the fact that both the initial and the final state have average and uncorrelated filling factors $\tfrac{1}{2}$. 


\end{document}